\documentclass[preprint,12pt]{elsarticle}

\usepackage{amssymb}
\usepackage{amsmath}
\usepackage{chemformula} % Formula subscripts using \ch{}
\usepackage[T1]{fontenc} % Use modern font encodings

\usepackage{amsmath}
\usepackage{amssymb}
\usepackage{soul}
\usepackage[margin=2.5cm]{geometry}
\usepackage{booktabs}   % For professional table formatting
\usepackage{siunitx}    % For aligning numbers
\usepackage{makecell}   % For multi-line cells
\usepackage{graphicx}   % For \resizebox (if needed)

\usepackage[hyphens]{url}

\usepackage{xcolor}
\definecolor{col_black}{RGB}{0,0,0}
\definecolor{col_review1}{RGB}{0,0,0} %{250,51,51} %{0,0,0} %   %\textcolor{review}{blabla...}
\definecolor{col_review2}{RGB}{0,0,0} %{51,51,250} %
\newcommand{\reviewfirst}[1]{\textcolor{col_review1}{#1}}
\newcommand{\reviewsecond}[1]{\textcolor{col_review2}{#1}}

\journal{Composites Part B: Engineering}

\begin{document}

\begin{frontmatter}
% \title{The Effects of Hydration on the Elastic Modulus of Graphene Polymer Nanocomposites Through Ensemble Based Atomistic Simulations}
%\title{Uncertainty Quantification of the Thermo-Elastic Properties of Hydrated Epoxy Polymers and Graphene Enhanced Nanocomposites with Ensemble Based Molecular Dynamics}

\title{\textbf{Thermo-elastic properties of hydrated epoxy-graphene nanocomposites from ensemble-based molecular dynamics simulations}}

\author[Rennes]{Maxime Vassaux}
\author[CCS]{Werner A. M{\"u}ller}
\author[CCS]{James L. Suter}
\author[Hexcel]{Alexandros Anastasiou}
\author[Hexcel]{Martin Simmons}
\author[Hexcel]{David Tilbrook}
\author[CCS,Advanced]{Peter V. Coveney}
\affiliation[Rennes]{organization={Univ. Rennes, CNRS, IPR (Institut de Physique de Rennes) - UMR 6251}, 
city={Rennes},
postcode={35000},
country={France}}
\affiliation[CCS]{organization={Centre for Computational Science - University College London},
addressline={20 Gordon Street}, 
city={London},
postcode={WC1H 0AJ},
country={United Kingdom}}
\affiliation[Advanced]{organization={Advanced Research Computing Centre - University College London},
addressline={20 Gordon Street}, 
city={London},
postcode={WC1H 0AJ},
country={United Kingdom}}
\affiliation[Hexcel]{organization={Hexcel Composites},
city={Cambridgeshire},
postcode={CB22 4QD},
country={United Kingdom}}

%% Abstract 250-word limit
\begin{abstract}
Epoxy-based materials are inherently hygroscopic, absorbing moisture from the environment, which can significantly alter their short and long-term performance. The presence of graphene is often considered as a potential candidate to act as a microscopic barrier, mitigating the adverse effects of hydration on the matrix. This study investigates the impact of hydration on the glass transition and elastic mechanical properties of epoxy resins and their graphene nanocomposites, focusing on water content up to $5$ \%wt. Using large-ensemble molecular dynamics simulations, we analyze the temperature-driven glass transition and mechanical response of both neat epoxy and epoxy-graphene systems under varying hydration levels. Our results reveal a distinct threshold at $3$ \%wt water content: below this, hydration primarily reduces the glass transition temperature, while mechanical properties remain unaffected. \reviewsecond{Beyond $2$ \%wt}, however, the mechanical properties deteriorate, highlighting a non-linear sensitivity to water uptake. Furthermore, we emphasize the critical role of ensemble size in ensuring the reliability of molecular dynamics predictions for such heterogeneous systems. Our simulations demonstrate that ensembles substantially larger than current state-of-the-art standards are necessary to achieve converged distributions of the predicted mechanical properties, particularly in highly heterogeneous hydrated epoxy-graphene nanocomposites. These findings provide novel insights into the hydration behavior of epoxy-based materials and underscore the \reviewfirst{limited} potential of graphene to enhance their environmental resistance. This work also advances the understanding of structure-property relationships in polymer nanocomposites, offering guidance for the design of more robust materials in humid environments.
\end{abstract}

\end{frontmatter}

\newpage

%%%%%%%%%%%%%%%%%%%%%%%%%%%%%%%%%%%%%%%%%%%%%%%%%%%%%%%%%%%%%%%%%%%%%
%% Introduction
%%%%%%%%%%%%%%%%%%%%%%%%%%%%%%%%%%%%%%%%%%%%%%%%%%%%%%%%%%%%%%%%%%%%%

\section{Introduction}

% Epoxy resins, mechanical properties and alteration by absorbed water
Epoxy resins are widely used in industries such as aerospace, automotive and microelectronics, owing to their superior thermal stability, mechanical strength, and favourable strength-to-weight ratio. The exposure of epoxy networks to humid or wet environments degrades their mechanical and thermal properties, including reductions in Young's modulus, tensile strength, bulk modulus, and glass transition temperature, accompanied by plasticization, swelling, microcracking, and chemical degradation \cite{morgan_effect_1980,tam_moisture_2015}. Indeed, their relatively high moisture permeability, compared to metals or ceramics, prevents the total exclusion of water from epoxy resins during or after their synthesis. However, the mechanisms by which water molecules alter the material properties of epoxy resins remain poorly understood.

% Former studies on the interaction between water and epoxy resins (exp., comp.)
The interaction between water molecules and epoxy resins has been the focus of several studies based on molecular dynamics (MD) simulations. Early on, the equilibrium organisation and dynamics of absorbed water was characterised by means of radial distribution functions (RDF) and mean squared displacements (MSD) of water molecules inserted in cured epoxy resins free volume \cite{seung_geol_lee_distribution_2010, yamamoto_dynamic_2021}. The dynamics of water molecules were found to be heterogeneous hinting that a distinction should be made between water molecules directly bound to the epoxy resin via hydrogen bonds and water molecules surrounded by other water molecules. In complementary FTIR spectroscopy and MD studies \cite{pandiyan_molecular_2015}, hydrogen bonding was also found to be a key metric to distinguish different types or dynamics of water molecules confined in epoxy resins, and helped to determine that water molecules tend to disperse in epoxies rather than forming aggregates. At higher water content, reaching up to $12$\%, however, cluster analysis revealed that water molecules finally tend to aggregate, with cluster size increasing exponentially with the water content  \cite{zhang_characteristics_2018}. This same study also hinted that at low hydration (\reviewsecond{below} $3$\%) an increase of Young's modulus could be found, while the same modulus started to decrease at higher hydration (\reviewsecond{above} $3$\%). An inverse trend has been found recently using reactive potentials, and low water hydration, approximately $1.5$\%, causing a decrease of the Young's modulus of $40$\% but an increase of the bulk modulus \cite{hou_molecular_2024}. Overall, although it has been the subject of several papers, the effect of hydration on the elastic properties has not fully been determined. In general terms, there is a lack of reproducibility in most reported MD studies, frequently with large and overlapping error bars \cite{frombgen_uncertainty_2026}. The case of MD simulations of complex systems such as hydrated epoxy resins is particularly striking, whereby the mechanical properties at different hydration levels cannot be differentiated \cite{sheng_molecular_2021}.

% Former studies involving molecular dynamics simulation of hydrated epoxy resins, and potentially involving graphene
Overall, this calls for new investigations using MD simulations, this time, with much larger ensembles. In order to enhance the moisture resistance of epoxy systems, the incorporation of nanofillers such as graphene has attracted considerable attention \cite{PERETZDAMARI2019105207}. Graphene offers outstanding mechanical strength and interfacial interactions with polymer matrices, making it a promising reinforcement for epoxy composites \cite{PAPAGEORGIOU201775}. Using large-scale large-ensemble MD simulations, we recently demonstrated the mechanisms enabling the reinforcement of polymers by graphene \cite{suter_large_2023}. When it comes to hydration, graphene is considered for multiple reasons: (i) its barrier effects, reducing water uptake, and (ii) to mitigate the decrease of mechanical properties of epoxy resins with increasing hydration. Early experimental results obtained with graphene-oxide supported such hopes \cite{Liu_Wang_Huang_2016}, nonetheless clear mechanistic explanations remained missing. This led to several studies of hydrated epoxy-graphene nanocomposites using MD simulations. These confirmed that in graphene-oxide systems, water diffusion was found to be lower due to its affinity with the reinforcement \cite{li_structures_2020}. The mechanical properties of graphene-epoxy nanocomposites were investigated observing the degradation with increasing hydration of the interface between the matrix and the nanoparticle \cite{kwon_molecular_2019, YANG2019425, zhang_effect_2023}. Moreover, these studies left bulk behaviour unexplored and focused on only two hydration levels.

% Objective of the paper.
In this work, we employ MD simulations to systematically investigate the effects of water content and graphene nanoparticle inclusion on the properties of water and thermomechanical properties of crosslinked epoxy networks. We focus on water structure, density, glass transition and elastic properties. We model the common and commercially available epoxy resin model epoxy diglycidyl ether of bisphenol F (DGEBF) cured with 4,4'-methylene-bis-(2,6-diethylaniline) (MDEA) and provide insight into the molecular mechanisms by which graphene reinforcement modifies the properties of hydrated epoxy nanocomposites.

To make actionable predictions from molecular dynamics (MD) simulations, we ensure control over the intrinsic uncertainty of our predictions. In MD simulations, the stress tensor values, and by extension, the elastic moduli, calculated from finite deformations have significantly high variance and act chaotically~\cite{wan_uncertainty_2021}. We have shown in our previous studies that ensemble MD, where numerous independent replicas of the simulation system are run with different starting conditions, provides the means to quantify the aleatoric uncertainty through ensemble averaging, thereby creating reproducible estimates with known errors \cite{vassaux_ensembles_2021}. We have applied such an ensemble-based approach to determine the glass transition of epoxy resins \cite{suter_rapid_2025}.

In this study, \reviewsecond{we rigorously determine the number of replicas required to produce estimates of the elastic moduli of the epoxy resin systems.} Indeed, we show that using a small number of replicas, or short simulation times, leads to unreproducible estimates of the macroscopic properties.

%%%%%%%%%%%%%%%%%%%%%%%%%%%%%%%%%%%%%%%%%%%%%%%%%%%%%%%%%%%%%%%%%%%%%
%% Methods
%%%%%%%%%%%%%%%%%%%%%%%%%%%%%%%%%%%%%%%%%%%%%%%%%%%%%%%%%%%%%%%%%%%%%

\section{Methods}

\subsection{Epoxy-graphene nanocomposites}

\begin{figure}
    \centering
    \includegraphics[width=0.9\linewidth]{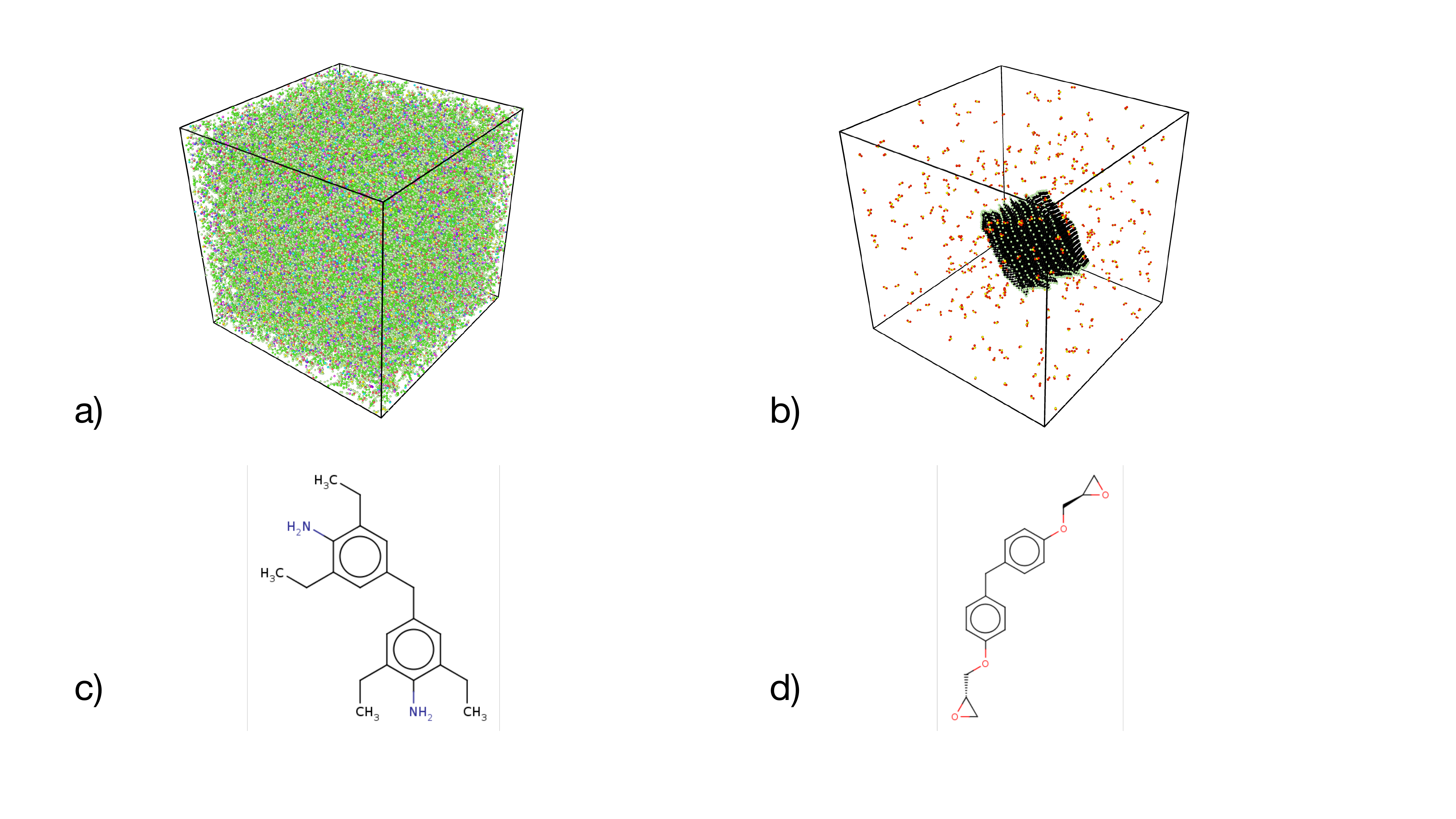}
    \caption{\textbf{Molecular models of epoxy-graphene nanocomposites.} (a) Visualisation of the hydrated epoxy-graphene nanocomposite molecular model. The molecular model is composed of $103,365$ atoms contained in a cubic box approximately $100$ \AA~wide at $300$ K. The model shown has a water content of $5$ \%wt. (b) Visualisation of the water molecules and the stack of graphene sheets in the molecular model shown in (a) as packed in the epoxy resin. (c,d) The epoxy resin is composed of a cross-linked network including the amine (c) 4,4'-methylene-bis-(2,6-diethylaniline) (MDEA) and the epoxy (d) diglycidyl ether of bisphenol F (DGEBF).
    % JLS: need colour codes for the atoms.
    }
    \label{fig:molecular_model}
\end{figure}

The investigated systems are cross-linked polymer chains consisting of the epoxy diglycidyl ether of bisphenol F (DGEBF) and the amine 4,4'-methylene-bis-(2,6-diethylaniline) (MDEA) (see figure \ref{fig:molecular_model}.a). The chemical structures of the two monomers can be found in figure~\ref{fig:molecular_model}.c,d. In addition to a system of pure epoxy, a pristine non-oxidised graphene flakes stack or tactoid is embedded inside the simulation box (see figure \ref{fig:molecular_model}.b). This graphene tactoid comprises $10$ layers of graphene with a radius of $25$ \AA. \reviewsecond{The description of the atomic and molecular composition of the simulated systems is summarised in table \ref{tab:composition}.}

\begin{table}[h]
\centering
\label{tab:composition}
\resizebox{\textwidth}{!}{%
\begin{tabular}{@{}l S[table-format=2.0] S[table-format=4.0] S[table-format=5.0] S[table-format=2.0] S[table-format=1.0] S[table-format=6.0] @{}}
\toprule
{\makecell[l]{Material}} &
{\makecell{Hydration \\ (\%)}} &
{\makecell{Number water \\ molecules}} &
{\makecell{Number \\ Graphene atoms}} &
{\makecell{Unreacted \\ DGEBF monomers}} &
{\makecell{Unreacted \\ MDEA monomers}} &
{\makecell{Atoms in \\ cross-linked network}} \\
\midrule
Epoxy & 0 & 0 & 0 & 24 & 3 & 95478 \\
Epoxy & 1 & 364 & 0 & 24 & 3 & 95478 \\
Epoxy & 2 & 735 & 0 & 24 & 3 & 95478 \\
Epoxy & 3 & 1114 & 0 & 24 & 3 & 95478 \\
Epoxy & 4 & 1501 & 0 & 24 & 3 & 95478 \\
Epoxy & 5 & 1895 & 0 & 24 & 3 & 95478 \\
\addlinespace[0.3cm]
Epoxy/Graphene & 0 & 0 & 6600 & 24 & 3 & 95478 \\
Epoxy/Graphene & 1 & 405 & 6600 & 24 & 3 & 95478 \\
Epoxy/Graphene & 2 & 818 & 6600 & 24 & 3 & 95478 \\
Epoxy/Graphene & 3 & 1239 & 6600 & 24 & 3 & 95478 \\
Epoxy/Graphene & 4 & 1669 & 6600 & 24 & 3 & 95478 \\
Epoxy/Graphene & 5 & 2108 & 6600 & 24 & 3 & 95478 \\
\bottomrule
\end{tabular}%
}
\caption{Atomic and molecular composition of the simulated systems.}
\end{table}

%This number of layer is set to match our reference experimental systems, described in table~\ref{tab:ex-results}, which consisted of between 10 and 20 layers of graphene. The size and number of layers of graphene were chosen to be representative of the experimental system. 
Due to the polymeric material being studied, the systems were parametrised with the class II polymer consistent force field (PCFF+)~\cite{sun_ab_1994,maple_derivation_1994,hwang_derivation_1994}. The pair-wise non-bonded interactions between the graphene sheets are also modelled using the all-atom PCFF+ parameters. It has been shown in prior studies that PCFF+ outperforms other all-atom force fields, such as CHARMM and Amber~\cite{bejagam_molecular_2020}, when calculating material properties such as the glass transition temperature.

The modelling of the epoxy-graphene nanocomposites is proceeded as follows. First, one graphene tactoid is randomly inserted inside an empty isotropic simulation box. Then, the epoxy and amine monomers are randomly packed around the graphene tactoid. 
%To match experiments, shown in Table~\ref{tab:ex-results}, the epoxy was packed so that the percentage by weight of the graphene would be $10$ \% of the system. 
%For this reason, due to the relatively small size of the simulated graphene flakes, with each flake being composed of around 650 atoms, the graphene was not oxidised at 2.8 \% like in the experiment, due to the small amount of oxygen that would have been simulated. % \mycomments{JLS: Be careful here - any reviewer might say that is not negligible, since GO can be that low, and that a small amount might significantly change properties. I'd say something about we want to examine the effect of the pristine graphene surface, which comprises 97.2\% of the experimental graphene.}
The epoxy amines were cross-linked using the MedeA~\cite{noauthor_materials_2024} thermoset builder software. \reviewsecond{The thermoset builder uses a starting capture sphere radius of $0.5$ nm, with maximum radius of search $15.0$ nm, and an increment of $0.01$ nm (maximum cycles limited to 800) through the iterative procedure.} The process of building the thermoset alternates between energy minimization, molecular dynamics and topology modifications forming new cross-links between the epoxy and amine species whenever their proximity satisfies a defined distance criterion. The dynamics simulation of the cross-linking of the monomers is computed \reviewsecond{at timesteps of $1$ fs via successive NVT ensembles of $10$ ps at $700$ K and NPT ensembles of $50$ ps at $700$ K and $1$ atmosphere} to ensure that the monomers were free to move and form the necessary bonds. The simulation was stopped when $90$ \% of the epoxy groups had formed bonds with the amines. Due to the likelihood of ring catenation, a higher bond conversion was not used. The pure epoxy system is obtained from the graphene systems by removing the graphene sheets and re-equilibrating the system. % Need to check this, why not just run the cross-linking without de graphene tactoid? 
After cross-linking, arriving at the final polymer, the system was then equilibrated at $700$ K by $1$ ns of NVT followed by $10$ ns of NPT at $1$ atm.

\subsection{Hydration}

Hydration is modelled inserting water molecules to achieve a given weight ratio (see figure \ref{fig:molecular_model}.b). Water molecules are modelled using the rigid SPC/E water model run under the SHAKE algorithm~\cite{ryckaert_numerical_1977}. Following the initial equilibration of the system, water molecules were inserted via random packing at $700$ K to achieve specified weight percentages (\%wt). \reviewsecond{The detailed description of the water content of the simulated systems is summarised in table \ref{tab:composition}.} To get a statistically robust sample of the epoxy and epoxy-graphene nanocomposite systems, for each water content weight, using different random seeds during the random insertion, ten replicas were produced where the water molecules were packed in different configurations, and the velocities of molecules were initialised with different random seeds drawn from a Maxwell-Boltzmann distribution~\cite{suter_rapid_2025}. 

\subsection{Glass transition}

Through simulated annealing, after the water molecules were inserted, a step-wise glass transition temperature  $T_g$ simulation was run from $700$ K to $300$ K. The $T_g$ simulation used a step size of $10$ K, where each step comprised $0.5$ ns of NVT followed by $2$ ns of NPT at $1$ atm. \reviewsecond{Such a quenching rate allows us to recover experimental values of the density of dry epoxy resins \cite{suter_rapid_2025}.}

Using atomistic simulations, polymers have long relaxation times, requiring long simulations of longer than $100$ ns of NPT to reach a state where the density reaches a maximum and is constant~\cite{suter_rapid_2025}. In addition to calculating the glass transition, the simulated annealing serves two purposes; not only does it lead to a series of ten statistically different starting configurations for the elastic simulations, but through a combined duration of $90$ ns of NPT, the simulated annealing also directly ends on a value for the density of the polymer at $300$ K close to equilibrium.

We have generated $10$ replica during the hydration simulation, this means that $10$ step-wise glass transition temperature simulations were performed for each \%wt, one for each water configuration.

The glass transition temperature is calculated through a hyperbola fitted across all density measurements, using the method of \citeauthor{patrone_uncertainty_2016}. The density is measured at $26$ different temperatures spanning from $700$ K to $300$ K. Each density at a given temperature is an ensemble average of $10$ replicas after $2$ ns of sampling time. Denoting $\rho$ to be the average density at a temperature point $T$,
\begin{equation}
    \rho (T) = \rho_0 - a \left(T - T_0 \right) - b \left(~ \frac{1}{2}\left(T - T_0 \right)+\sqrt{\frac{\left(T-T_0\right)^2}{4}+e^c} ~\right)
\label{patrone}
\end{equation}
where $T_0,\rho_0,a,b,c$ are positive real constants and $e$ is Euler's constant. The glass transition temperature is given by $T_0$, representing the hyperbola's centre. The hyperbola is fit using the least squares algorithm.

% \mycomments{JLS: you should also say something here that this procedure, as well as providing the Tg, is an effective way of annealing the system to 300K. A reviewer will ask you questions about how you determined you were in (quasi)-equilbrium at 300K. You mentioned the density - but what do you mean by realistic? Due to the very long relaxation times you are very unlikely to exactly recreate the experimental densities.}

% To get a statistically robust sample of the epoxy and graphene nanocomposite systems, for each water content weight, using different random seeds during the random insertion, ten replicas were produced where the water molecules were packed in different configurations, and the velocities of molecules were initialised with different random seeds drawn from a Maxwell-Boltzmann distribution~\cite{suter_rapid_2025}. This meant that ten step-wise glass transition temperature simulations were performed for each wt\%, one for each water configuration. % This further implies that for each water weight, the starting systems for the elastic constant simulations were equilibrated independently.

\subsection{Mechanical deformation}\label{ssec:mech_methods}

We calculate the elastic constant through finite deformations at a fixed temperature, averaged from deformations carried out positively and negatively in six directions. Let $C_{jl}\in \mathbb{R}^{6\times6}$ be the 36-valued elastic constant tensor, then we calculate the modulus as follows:
\begin{align}
T_{11} &= \frac{C_{11}+C_{22}+C_{33}}{3}, \quad T_{12} = \frac{C_{12}+C_{13}+C_{23}}{3}\\
K &= \tfrac{1}{3}(T_{11}+2T_{12}) \\
G_1 &= \tfrac{1}{3}(C_{44}+C_{55}+C_{66})\\
G_2 &= \tfrac{1}{2}(T_{11}-T_{12}) \\
\nu &= (1+T_{11}/T_{12})^{-1} \\
E &= 3K(1-2\nu)
\end{align}
where $K$ is the bulk modulus, $G_1$ and $G_2$ are the shear modulus, $\nu$ is the Poisson's ratio, and $E$ is the Young's modulus. The values $T_{11}$ and $T_{12}$ are intermediate values to aid in the calculation of the moduli. Due to the symmetry of the elastic constant, we average the non-diagonal component pairs of the elastic constant tensor when calculating the moduli.

The elastic deformation simulations are performed under the NVT ensemble at $300$ K with a time step of $1$ fs. A strain of amplitude $1$ \% is applied over $10$ ns in the different directions, in compression and tension, and the resulting stress is sampled over $1$ ns. \reviewfirst{The resulting strain rate is $1$ $\mu$s$^{-1}$.} The validation of the mechanical deformation simulation protocol is supported by the supplementary figure $2$ \reviewsecond{which shows the evolution of the Young's modulus estimate with the duration of the deformation application and the size of the deformation amplitude. The Young's modulus estimations converge for deformation durations beyond $8$ ns.}

\subsection{Ensemble-based simulations}

\begin{figure}
    \centering
    \includegraphics[width=0.95\linewidth]{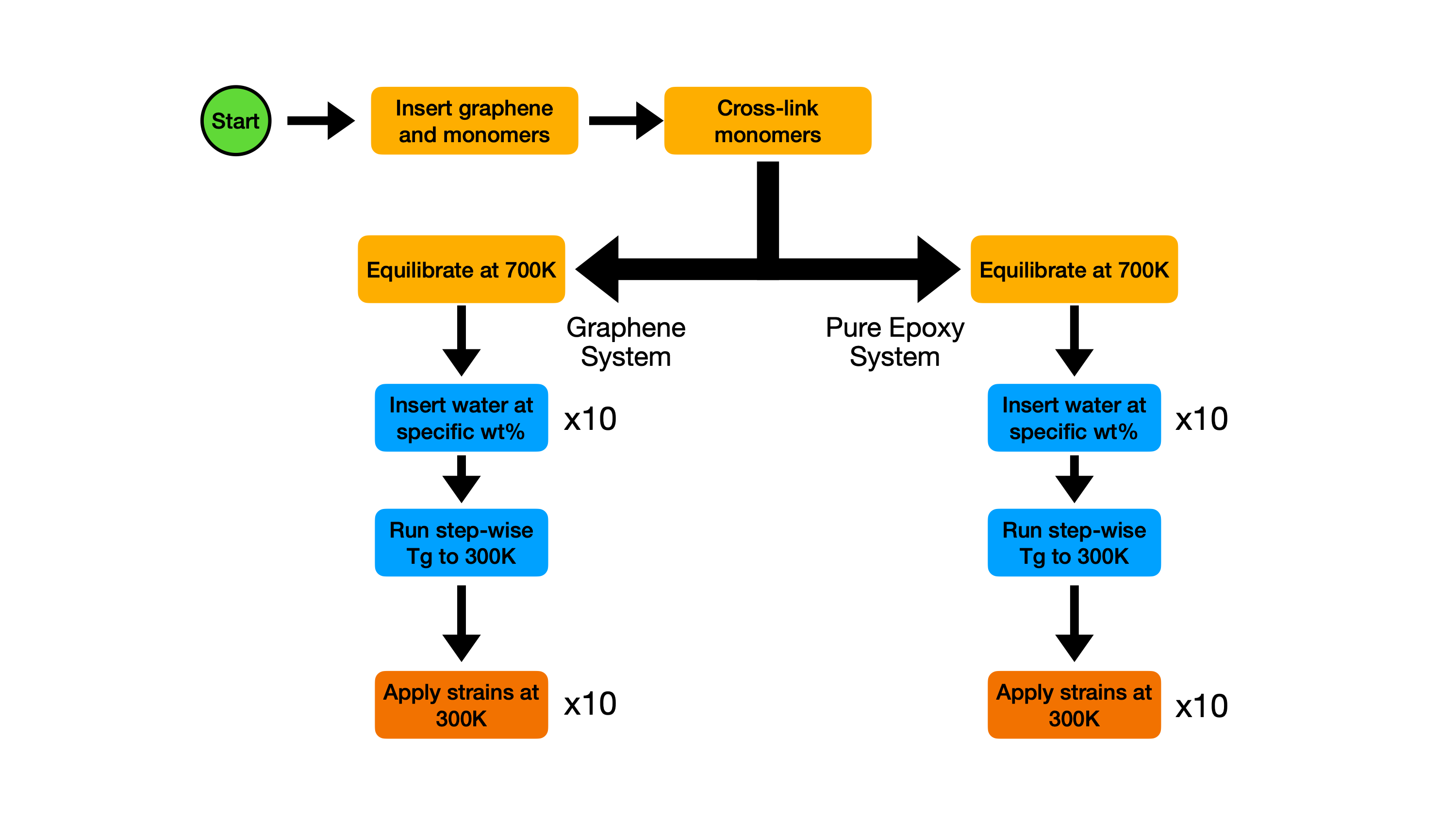}
    \caption{\textbf{Methodology for the simulation of synthesis and characterisation epoxy-graphene nanocomposites.} (i) The synthesis consists in a first cross-linking step of the epoxy monomers, in presence or not of a graphene tactoid. (ii) The cross-linked molecular models are equilibrated at $700$ K and $1$ atm. (iii) The molecular models are hydrated to the targeted hydration level. (iv) The hydrated molecular models are cooled down to $300$ K, during which $T_g$ is computed. (v) Last, the molecular models are strained in order to calculate the elastic constants. During synthesis and equilibration a single replica is simulated, during hydration and glass transition $10$ replica are simulated, last during mechanical deformation $10$ replica are simulated.
    %\mycomments{JLS: To make this clearer, I'd use a fan-out diagram for the replicas, i.e. show that from the equlibriated structures you made ten replicas of different water packing (replicas(wp)) (fan out 1), for each water content (fan out 2), and from each of these 100 replicas of elastic tensor computation(replicas(et)) (fan out (3)). Just use dots to indicate there are 10 / 100 replicas etc. You must emphasise that you have performed a VAST number of simulations here. }\wcomments{I'll remake the figure.}
    }
    \label{fig:methods}
\end{figure}

In short, $10$ replicas at the hydration stage, each replicated $10$ times at the mechanical deformation stage (see figure \ref{fig:methods}).

The velocities of the system are then reinitialised at $300$ K with different random seeds to generate $10$ further replicas for that water configuration. This leads to a total of $100$ different replicas for each material.

%%%%%%%%%%%%%%%%%%%%%%%%%%%%%%%%%%%%%%%%%%%%%%%%%%%%%%%%%%%%%%%%%%%%%
%% Results
%%%%%%%%%%%%%%%%%%%%%%%%%%%%%%%%%%%%%%%%%%%%%%%%%%%%%%%%%%%%%%%%%%%%%

\section{Results and discussions}

\subsection{Molecular structure of water}\label{ssec:water_structure}

\begin{figure}[h]
  \includegraphics[width=\textwidth]{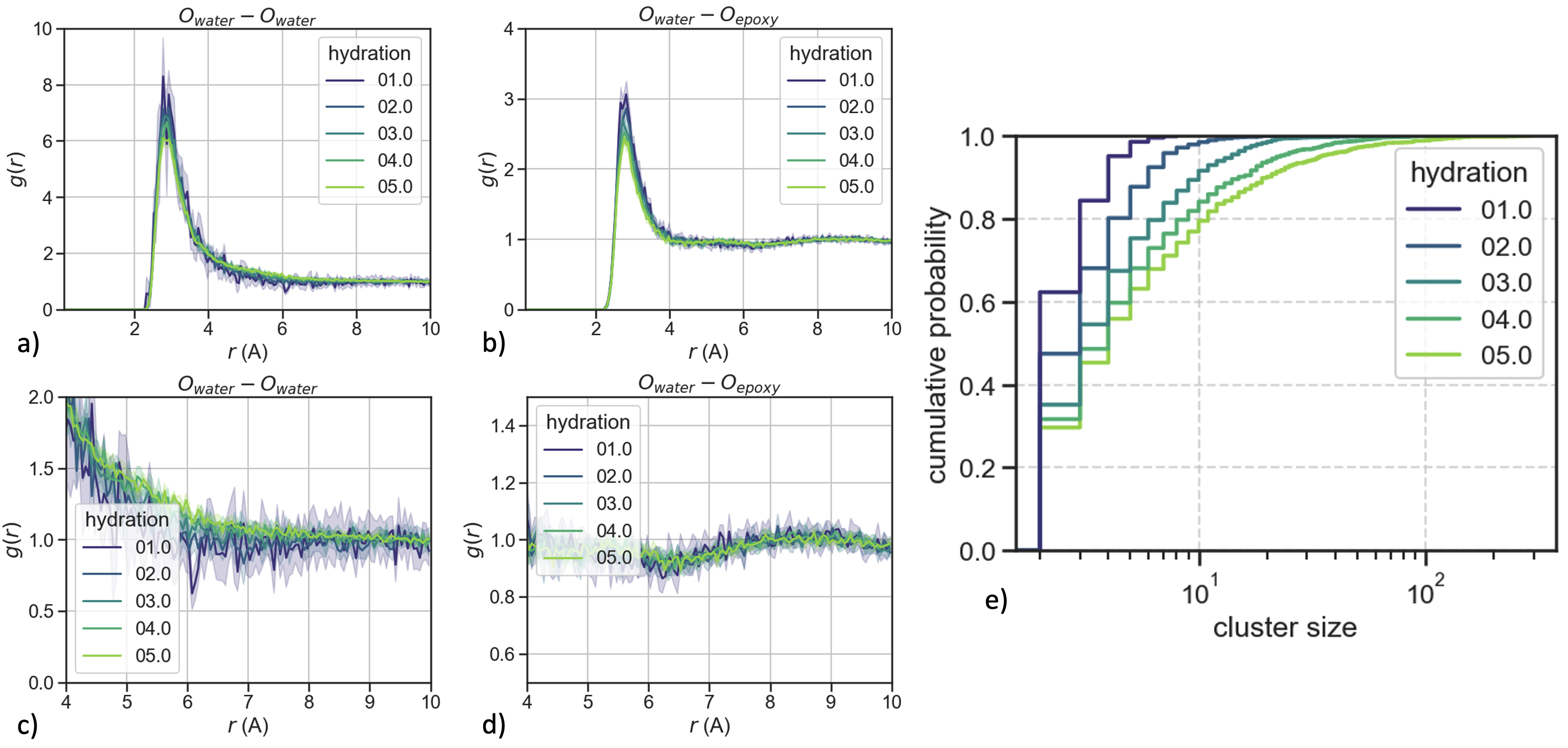}
  \caption{\textbf{Water structure in epoxy resins.} Water short-range and medium-range order in epoxy resins. (a-d) Evolution of the radial distribution functions (RDF) of the water and epoxy oxygen atoms with water content. (a) Water-water oxygen RDF up to $10$ \AA, (c) zoom on the $4$ to $10$ \AA~region. (b) Water-epoxy oxygen RDF up to $10$ \AA, (d) zoom on the $4$ to $10$ \AA~region. (e) Evolution of the distribution of the water molecules clusters sizes with the water content.}
  \label{fig:water_structure_resin}
\end{figure}

We investigate the structure of water molecules in the pure epoxy resin two ways (see figure \ref{fig:water_structure_resin}). First, by means of the partial radial distribution functions (RDF) of the oxygen atoms in water molecules with respect to the oxygen atoms in water molecules or in epoxy groups. Second, by means of a water molecules cluster size analysis. All RDF become smoother with increasing hydration, as statistics improve with the increasing number of water molecules. The RDF of water molecules oxygen atom with themselves shows a sharp first peak around $2.8$ \AA~which position remains stable with hydration. The height of the peak, however, decreases, and the right side of the peak tend to decrease slower with the distance between oxygen atoms $r$. As water molecules are being added, second and third peak, respectively at $4.5$ \AA~and $6-7$ \AA~in bulk water, appear. With increasing hydration, water molecules therefore start to aggregate and start to exhibit structural properties from bulk water. The RDF of water molecules oxygen and epoxy groups oxygen displays a similar trend, the first peak remains at $2.8$ \AA~and its intensity decreases with hydration, yet, at longer distances no changes are observed with increasing hydration.

The cluster size distribution (see figure \ref{fig:water_structure_resin}) reveals predominantly small water clusters (less than $10$ molecules) at low hydration ($1$ \%wt). Water is mostly bound to high-affinity polymer sites. At higher hydration, much larger water clusters are found as the high-affinity polymer site become saturated with water, and interwater interactions begin to dominate. As described earlier, the observed cross-over from predominately small, dispersed water clusters at low water uptake to larger clusters can be explained by competing affinity and geometric effects. At low water concentrations, water molecules are efficiently solvated by polar sites in the cross-linked epoxy (especially the hydroxyl groups), forming small hydrogen bonded complexes. As these sites approach saturation, additional water must occupy other environments and the water clusters grow. 

\begin{figure}[h]
  \centering
  (a)
  \includegraphics[width=.3\textwidth]{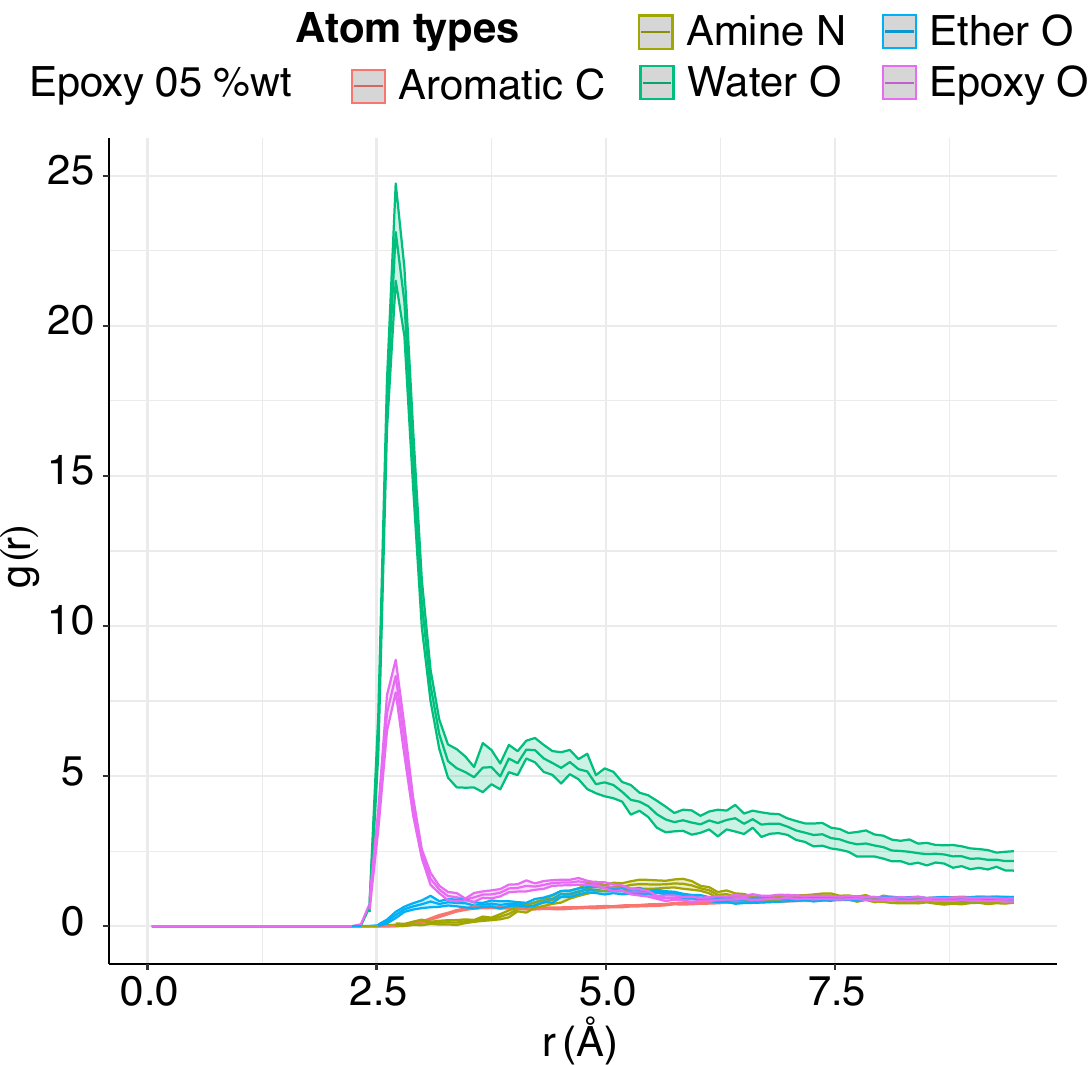}
  (b)
  \includegraphics[width=.3\textwidth]{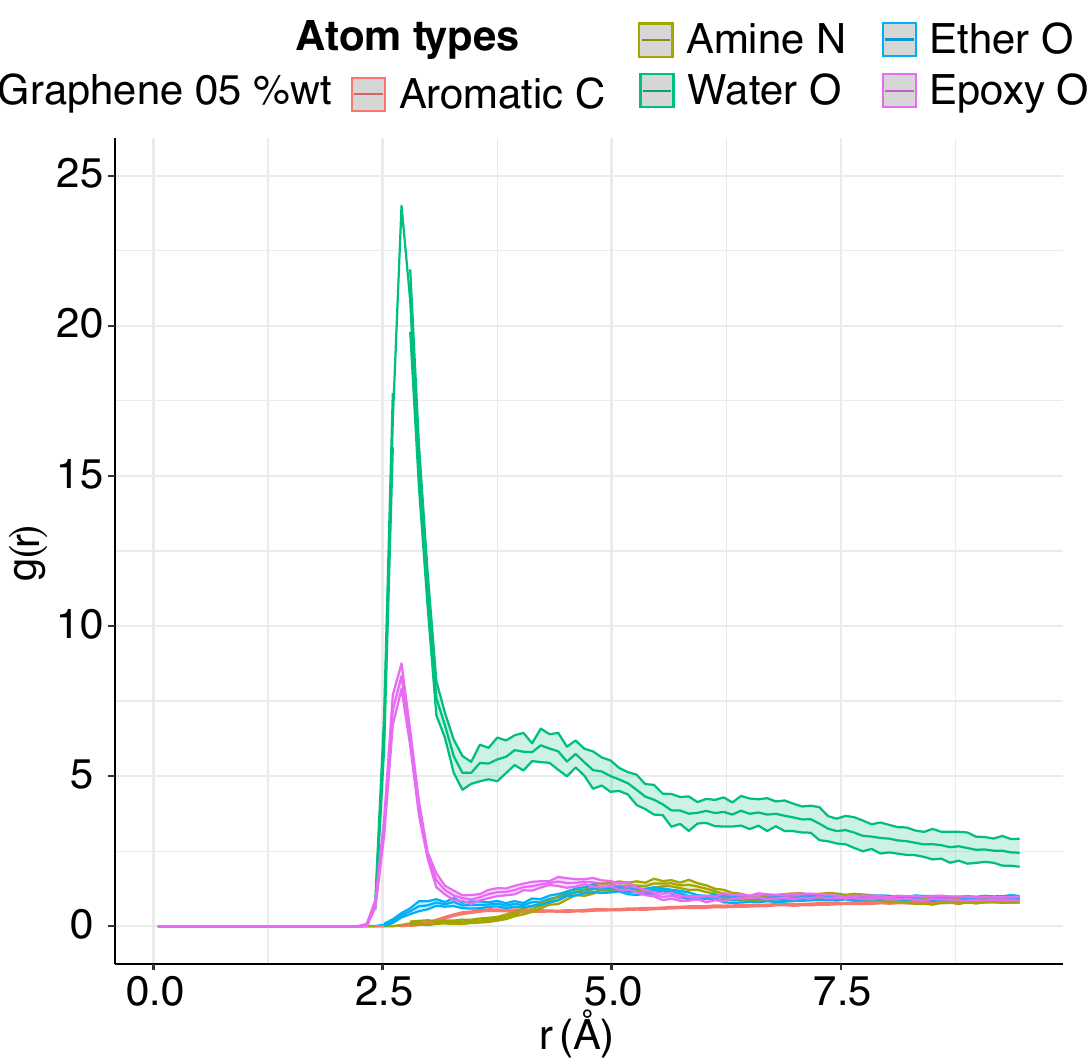}
  (c)
  \includegraphics[width=0.25\textwidth]{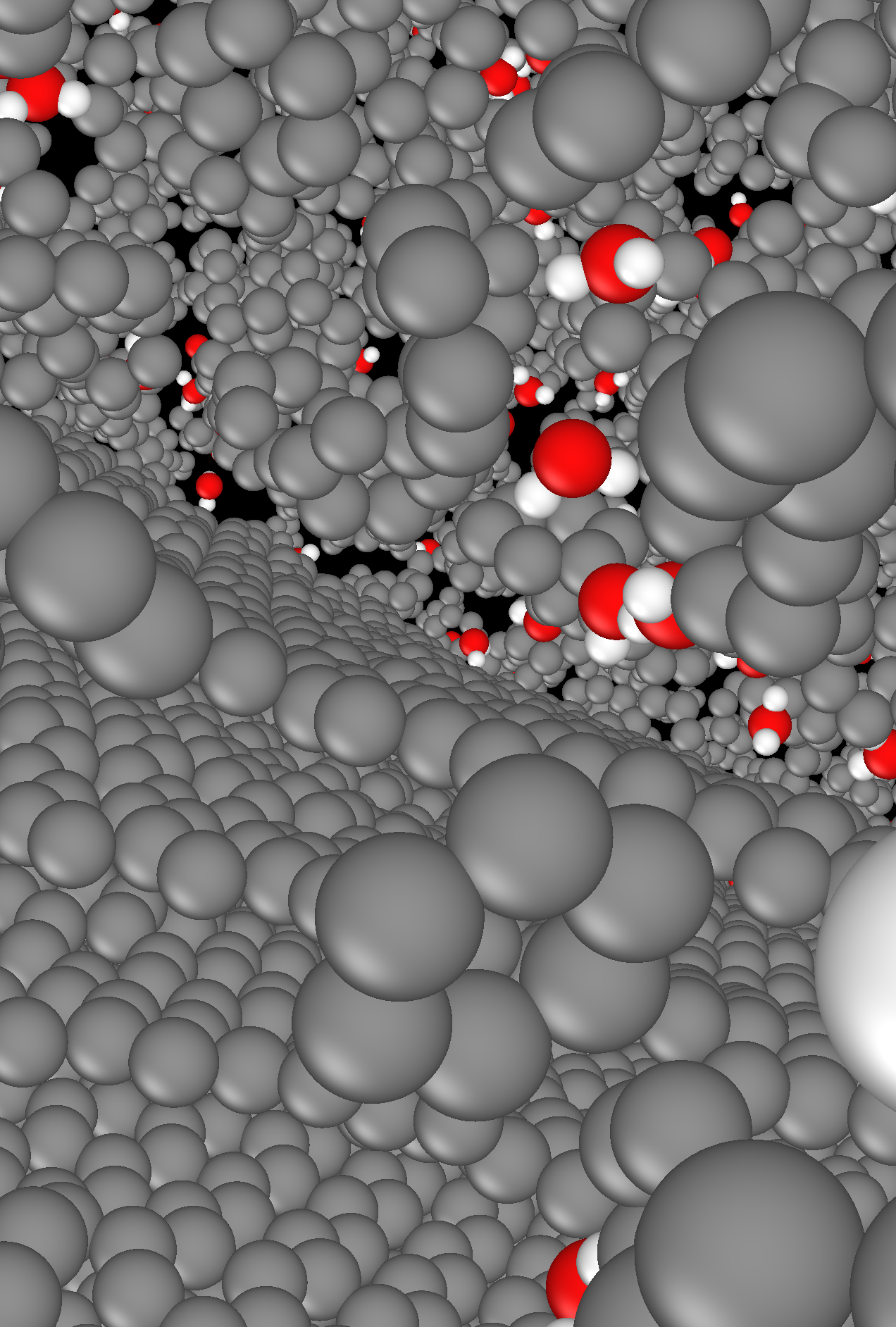}
  \\
  (d)
  \includegraphics[width=0.7\textwidth]{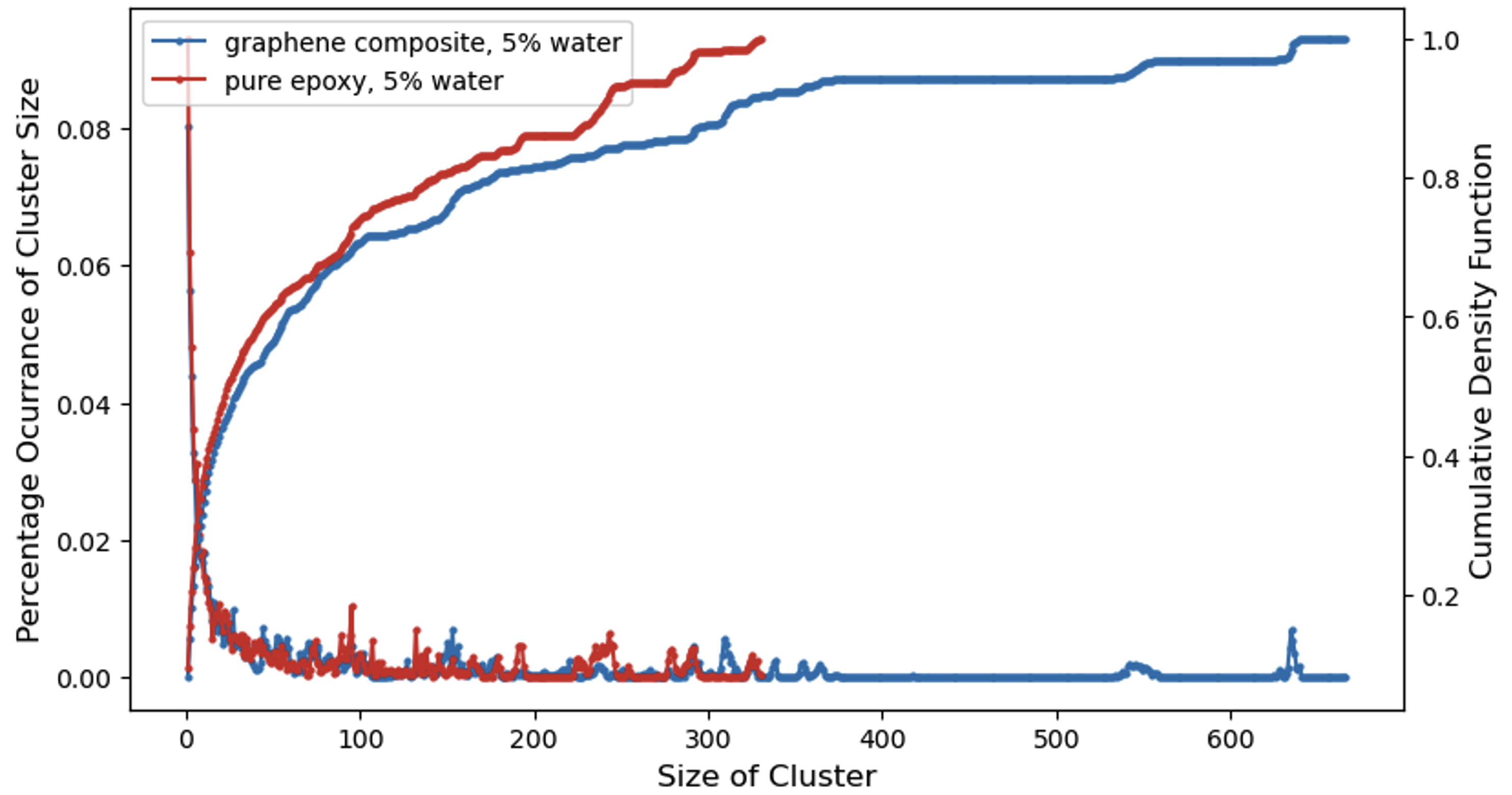}
  \caption{\textbf{Graphene-induced structural changes in water.} Water short-range and medium-range order at $5$\% hydration. Evolution of the RDF of between the water oxygen atoms and various epoxy chains atoms types: (a) in pure epoxy resins and (b) in graphene-epoxy nanocomposites. (c) \reviewfirst{Snapshot of the structure of an epoxy-graphene nanocomposite model at $5$\% hydration, with a focus on the interface of the graphene tactoid illustrating the absence of water molecules. (d)} Water molecule cluster size analysis (probability distribution and cumulative distribution) for the resin (in red) and the nanocomposite (in blue).}
  \label{fig:water_structure_graphene}
\end{figure}

We compare the water molecular structure at $5$\% hydration between the pure epoxy resin and the graphene-epoxy nanocomposite (see figure \ref{fig:water_structure_graphene}). RDF between water molecules oxygen atoms and a variety of different atom types in the epoxy chains are not affected by the insertion of graphene. The water molecules are primarily hydrogen bonded to other water molecules or the epoxy groups on the polymer chains. This demonstrates that the presence of graphene does not affect the local hydration environment around the polar oxygen atoms of the epoxy polymer, nor that, as expected, the water molecules do not reside near the graphene sheets. \reviewsecond{Indeed, no water molecule can be found less than $2.7$ \AA~away from the graphene sheets (see supplementary figure $1$.}

On the contrary, longer-range order observations, from the cluster size analysis, display a clear divergence, as the graphene containing systems contain a higher fraction of larger water clusters compared to the pure epoxy. We hypothesise that, although the graphene tactoid only occupies a relatively small volume, its high surface to volume ratio decreases the area available for water dispersion, thereby increasing the water-water interactions. \reviewfirst{This is supported by the observation of the interface region between the graphene and the epoxy resin at high hydration ($5$\%wt) which is void of water molecules.}

\subsection{Density and glass transition}

\begin{figure}[h!]
    \centering
    (a)
    \includegraphics[width=0.35\linewidth]{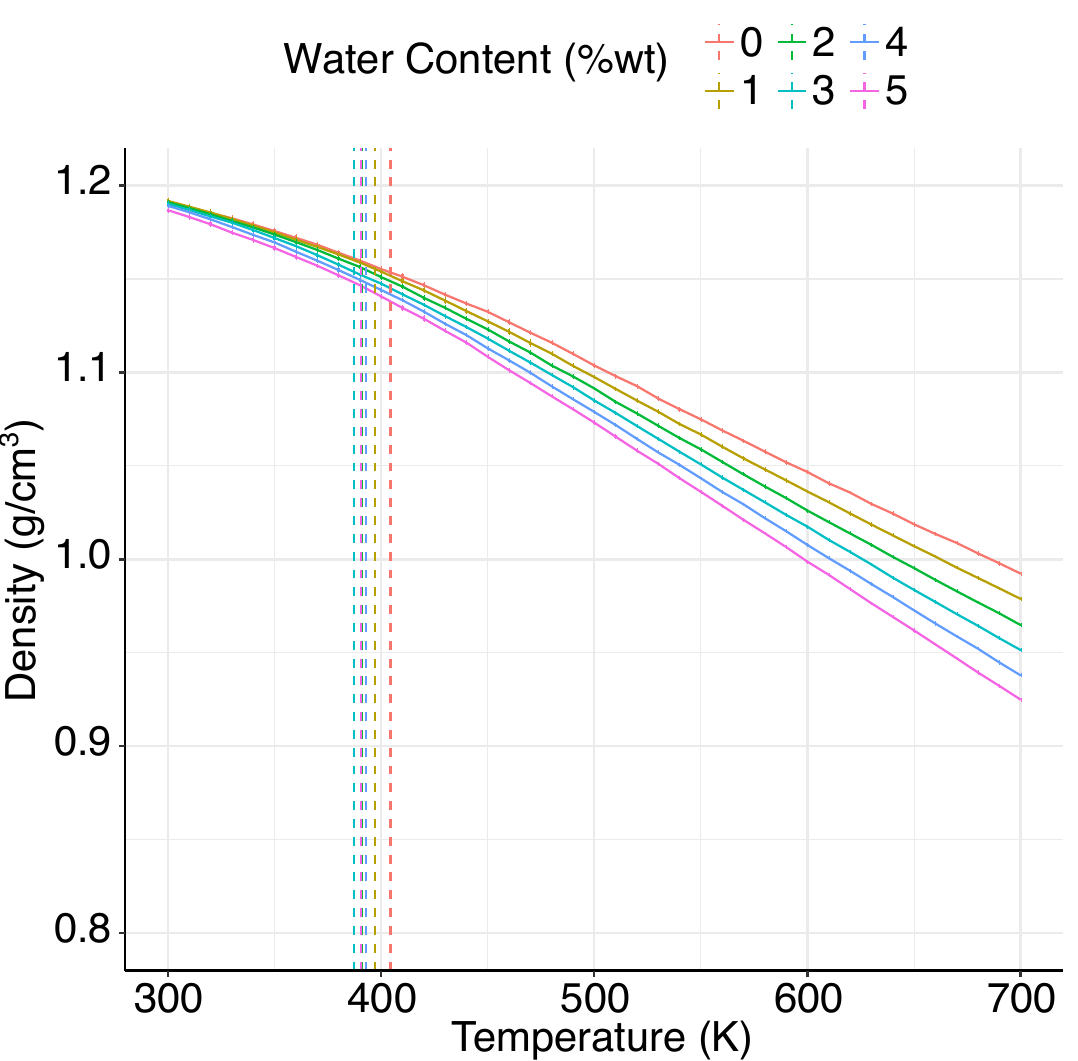}
    (b)
    \includegraphics[width=0.35\linewidth]{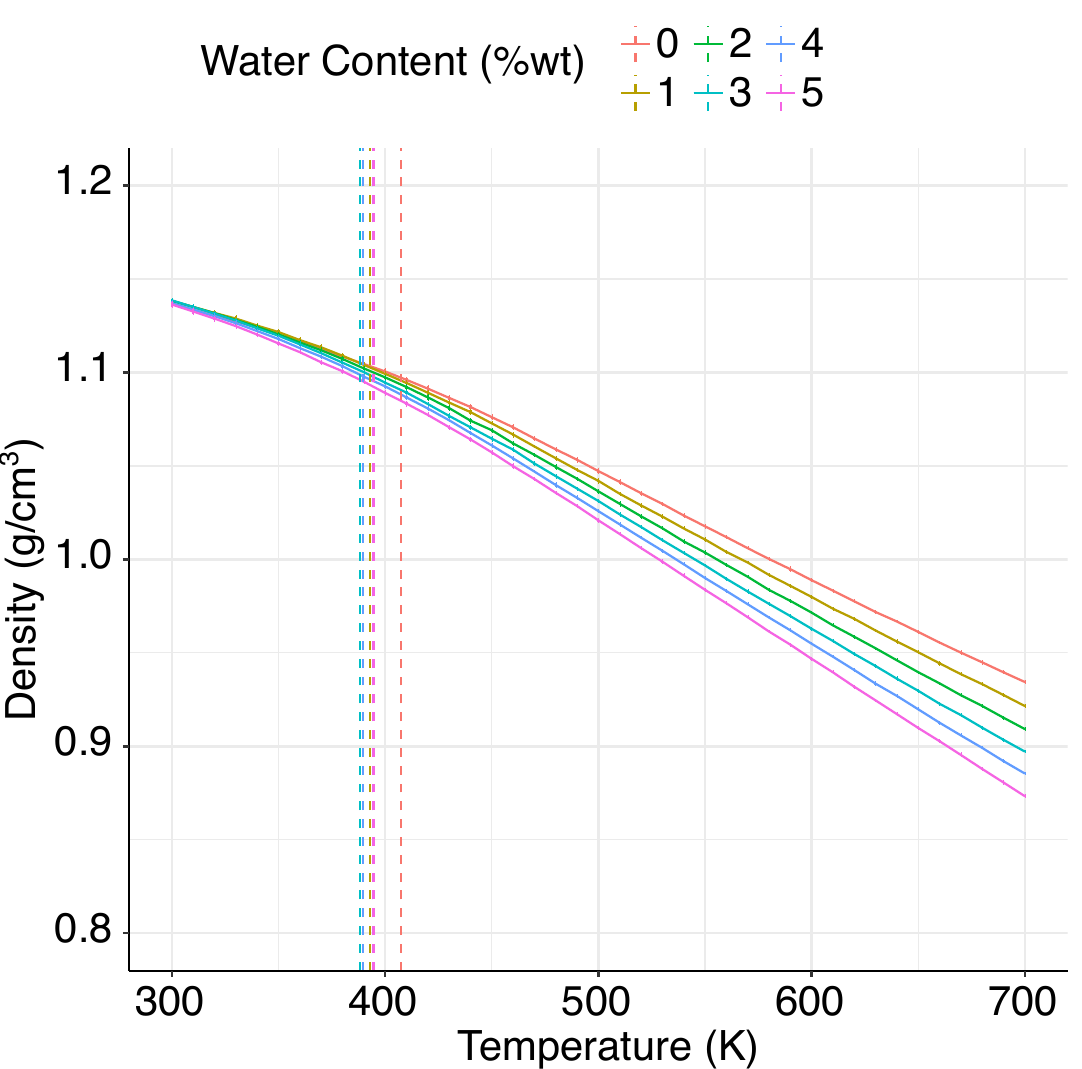}\\
    (c)
    \includegraphics[width=0.75\linewidth]{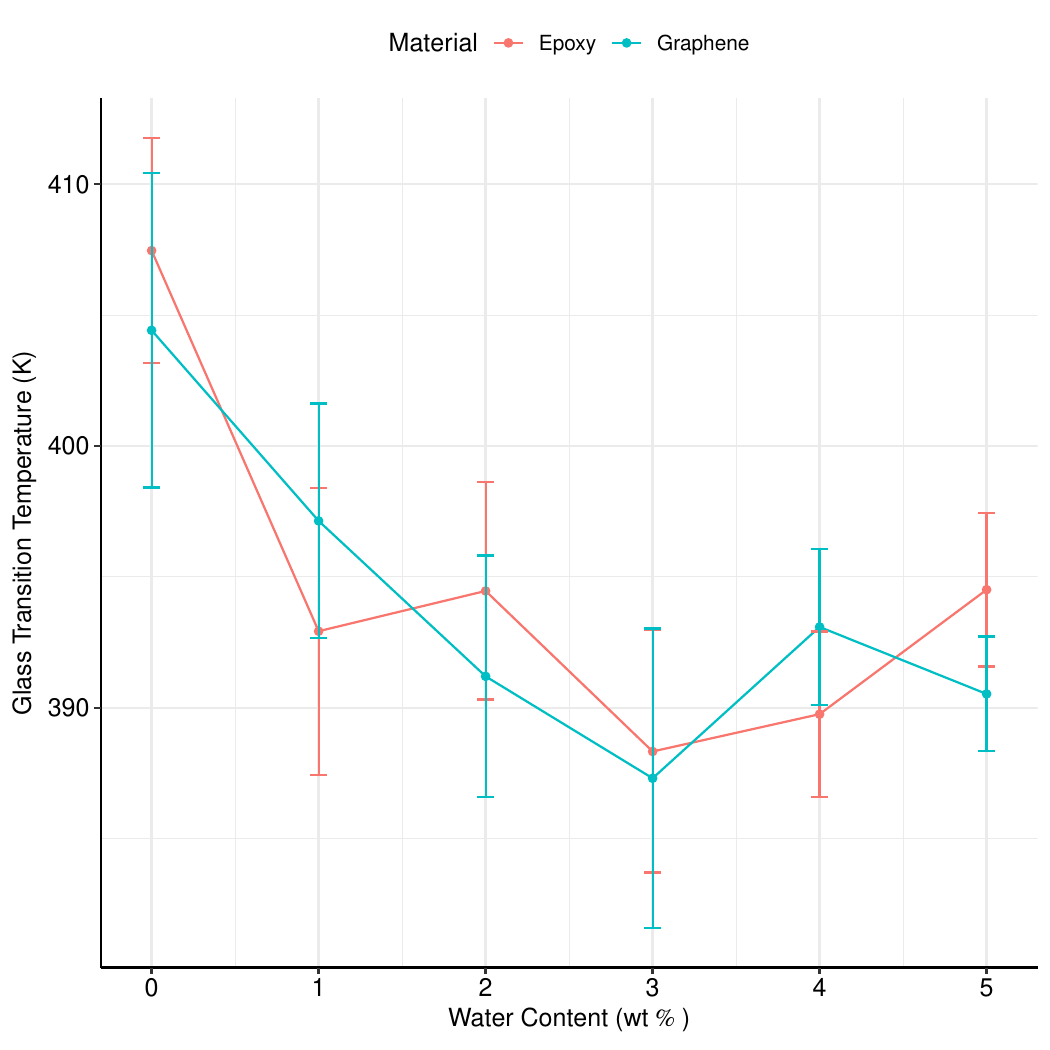}
    \caption{\textbf{Density and glass transition.} Influence of water content on the evolution of the (a) polymer and (b) graphene nanocomposite density with temperature. The vertical dashed lines label the glass transition temperature. (c) Evolution of the glass transition temperature with the water content. \reviewsecond{In (a) and (b), error bars correspond to the total standard deviation of all $10$ replicas at a given water content; in (c) error bars correspond to the regression error in the least squares fit used to optimise the glass transition temperature parameter.}}
    \label{fig:water-density-temp}
\end{figure}

We first focus on the glass transition phenomena, which we investigate observing the evolution of the density with temperature in our molecular models (see figure \ref{fig:water-density-temp}). In all molecular models, the density decreases nonlinearly with the temperature, displaying an inflexion point associated with the glass transition. Note that the standard error of density measurements over the $10$ replicas is displayed in figures \ref{fig:water-density-temp}.a,b, but is too small to be visible. The presence of water inside the epoxy resins and the epoxy-graphene nanocomposites systematically decreases the density across the whole temperature range. This is emphasised at higher temperatures. Meanwhile, the addition of graphene tends to decrease the density of the material, independently of the water content.

The hyperbola fitting of the evolution of the density with temperature (see equation \ref{patrone}) provides us with the glass transition temperature at different water contents (see figure \ref{fig:water-density-temp}.c). The error bars refer to the \reviewsecond{regression} error in the least squares fit used to optimise the glass transition temperature parameter, $T_0$, from the hyperbola in equation~\ref{patrone}. We compute glass transition temperatures in agreement with experimental data for pure dry epoxy resins, around $410$ K \cite{ribeiro_glass_2013, sharp_effects_2017}. The increase of the water content up to $3$\% reduces the transition temperature of epoxy resins down to $380$ K, then the transition temperature remains stable up to $5$\% water content. \reviewsecond{These results on the effect of hydration are consistent with earlier experimental and computational studies in predicting a decrease of the glass transition temperature \cite{sharp_effects_2017, yamamoto_dynamic_2021}. Further our predictions quantitatively agree best with the experimental data which is provided up to $2$ \% hydration \cite{sharp_effects_2017}.} Regarding epoxy-graphene nanocomposites, identical variations are observed. The inclusion of the graphene tactoid inside the molecular model of the epoxy resin does not modify the glass transition temperature.

% Discussion vs. former studies, vs. what is expected
We used a tactoid of ten sheets in our simulation to match the number of graphene sheets typically found in commercially available graphene. Our results show that the low interfacial area associated with this aggregated state and the lack of interaction between the graphene and the epoxy chains results in no discernable change in the glass transition temperature. Furthermore, the small difference in predicted glass transition temperatures between the epoxy polymers and graphene nanocomposites aligns well with results from comparable computational studies~\cite{shiu_characterizing_2014} of single replica simulations of atomistic dry epoxy polymer and dry graphene nanocomposite. Curiously, the absence of any effect from the insertion of graphene is contrary to what is expected. However, the potentially reduced mobility of the cross-linked epoxy chains in the immobilised interphase region near the graphene surface is indeed expected to yield an increase in the glass transition temperature.

Considering the effect of water molecules, we explain the decrease in glass transition temperature with increased mobility induced by the presence of water molecules. Firstly, penetration of water molecules into the epoxy matrix creates greater free volume for the epoxy chains to move. In addition, water molecules can compete with inter- and intra-chain hydrogen bonds, which will also allow the epoxy chains to move more freely. Since the glass transition temperature plateaus at $3$\% water, the interrupting of hydrogen bonding within the epoxy chains is likely to be the dominant factor in the reduction at low water contents. At $3$\% water content, the water molecules change from being mostly bound to the epoxy polar sites to becoming free water, thereby reducing the effective amount of chain-solvating water (see section \ref{ssec:water_structure}), therefore additional water does not increase chain mobility. If the decrease in glass transition temperature was dominated by additional free-volume, the glass transition temperature would be expected to continue decreasing past $3$\% water content as free-volume increases. 

\subsection{Mechanical properties}

\begin{figure}[h]
  (a)
    \includegraphics[width=.4\textwidth]{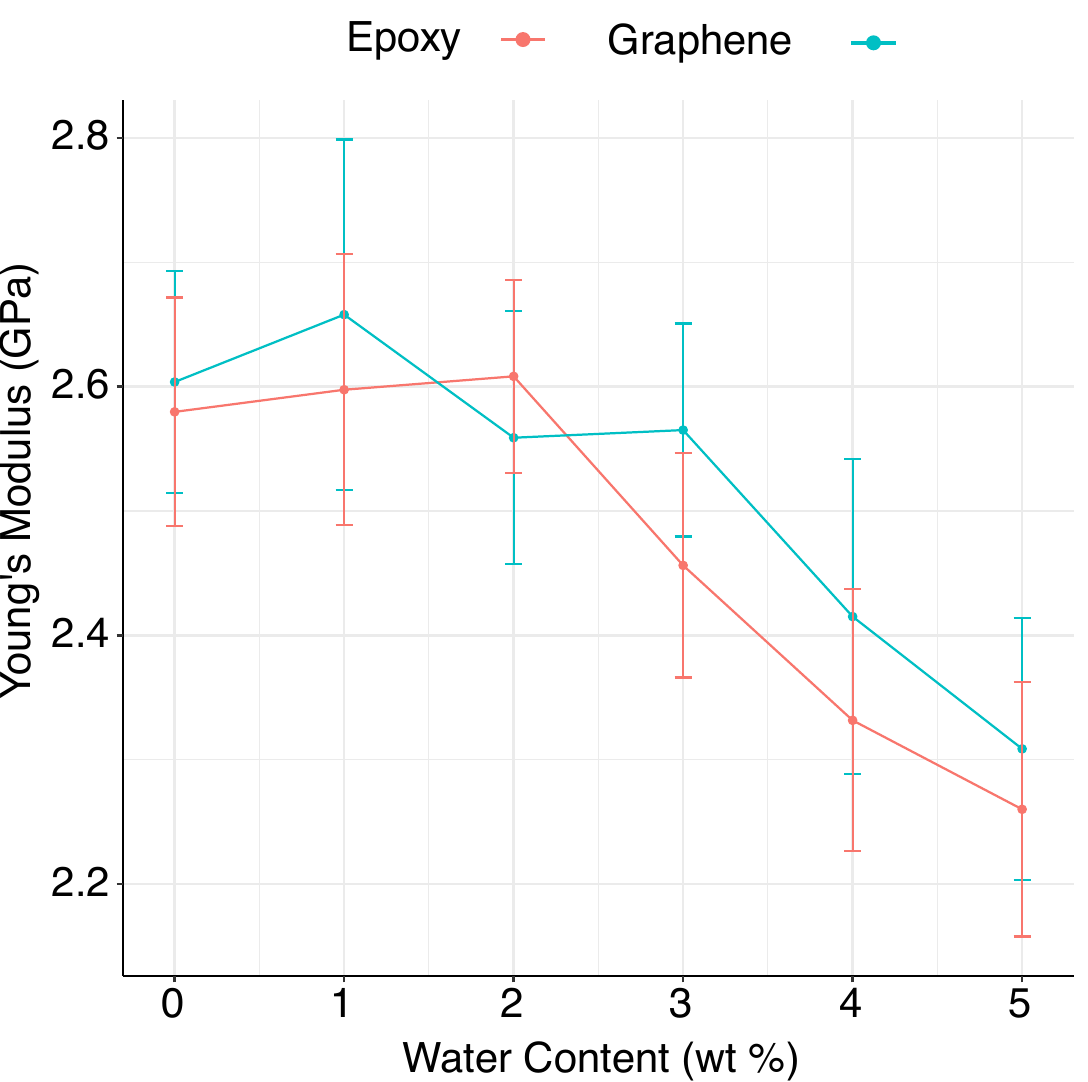}
  (b)
  \includegraphics[width=.4\textwidth]{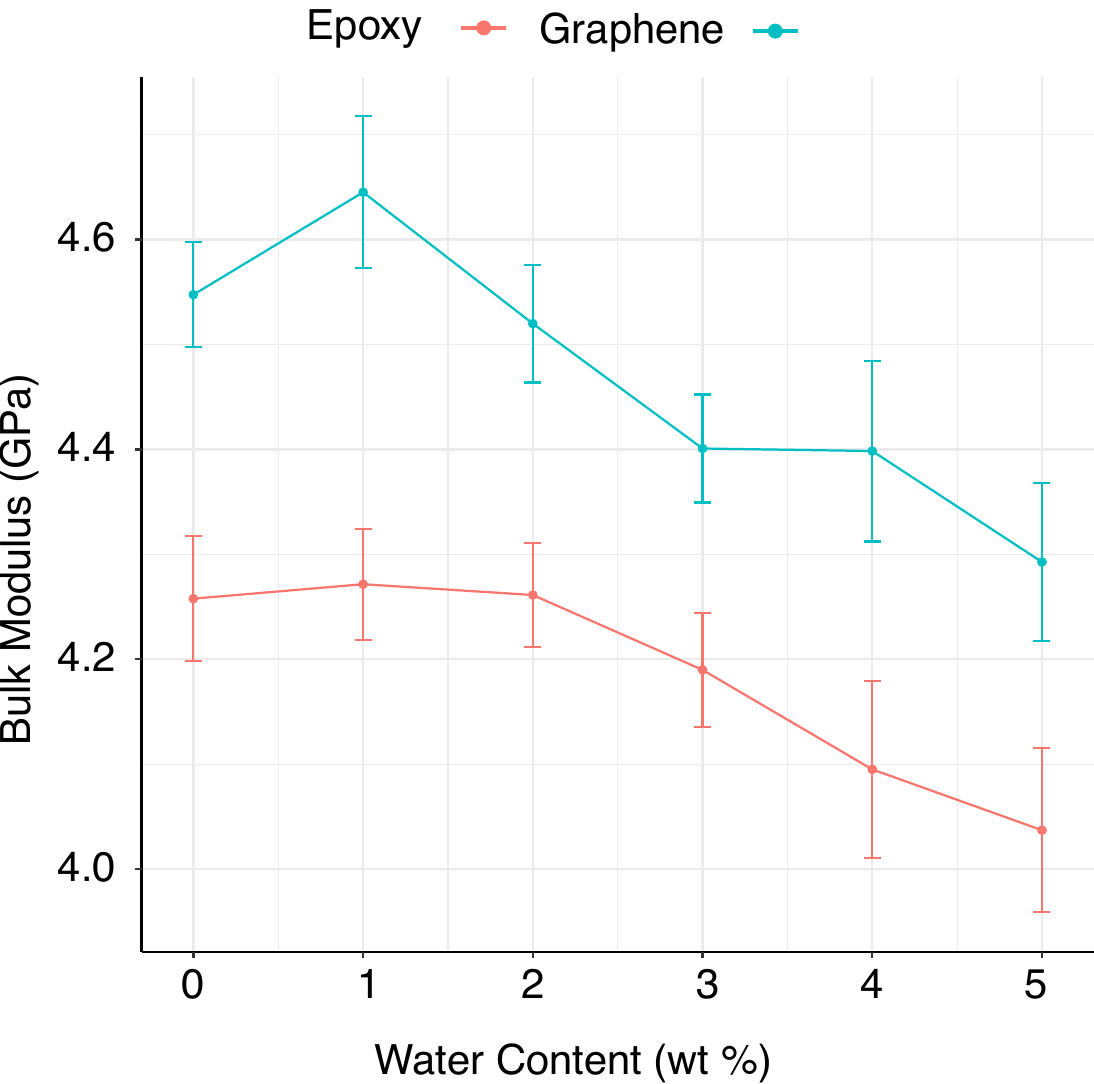}\\
  (c)
  \includegraphics[width=.25\textwidth]{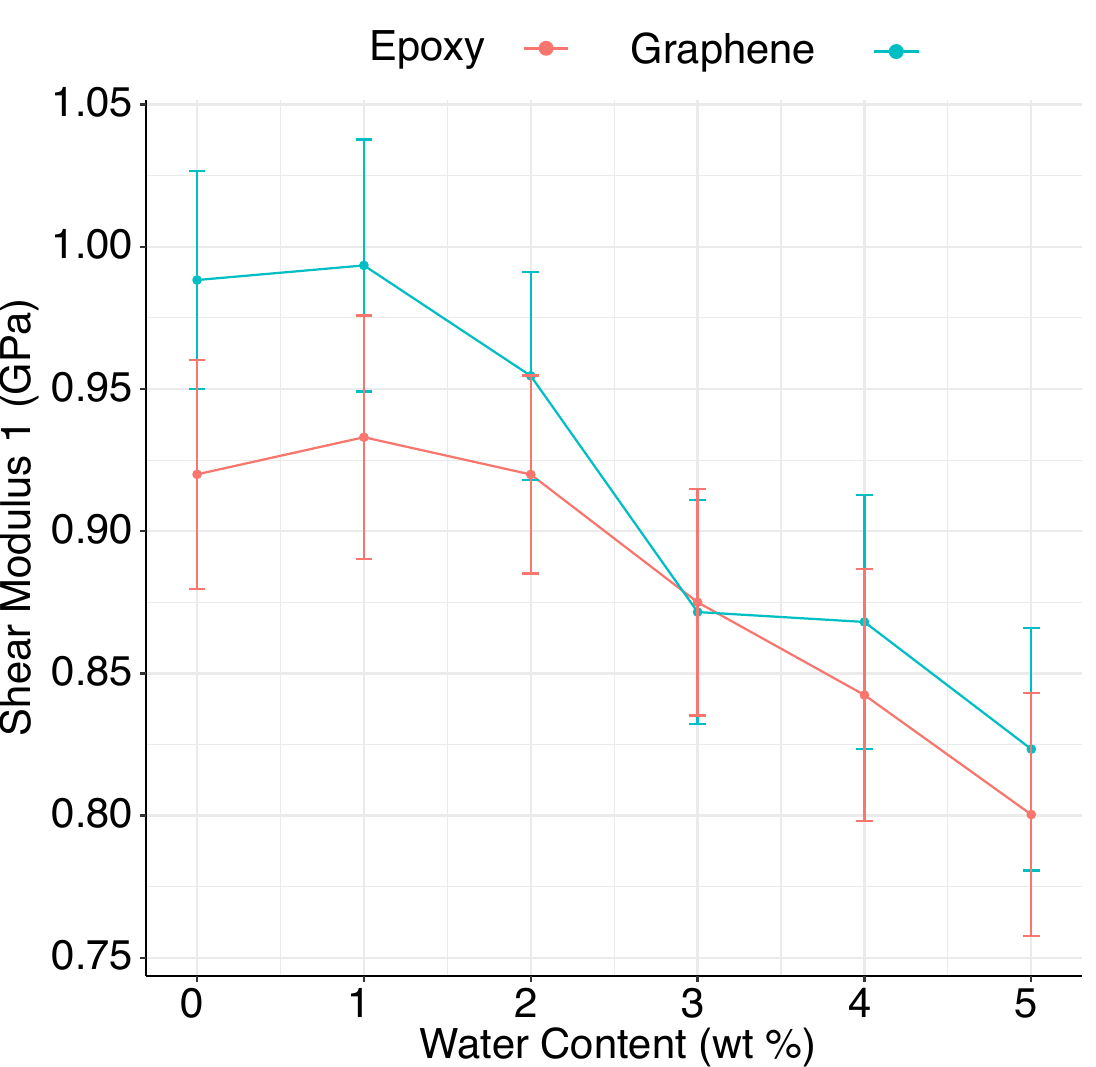}
  (d)
  \includegraphics[width=.25\textwidth]{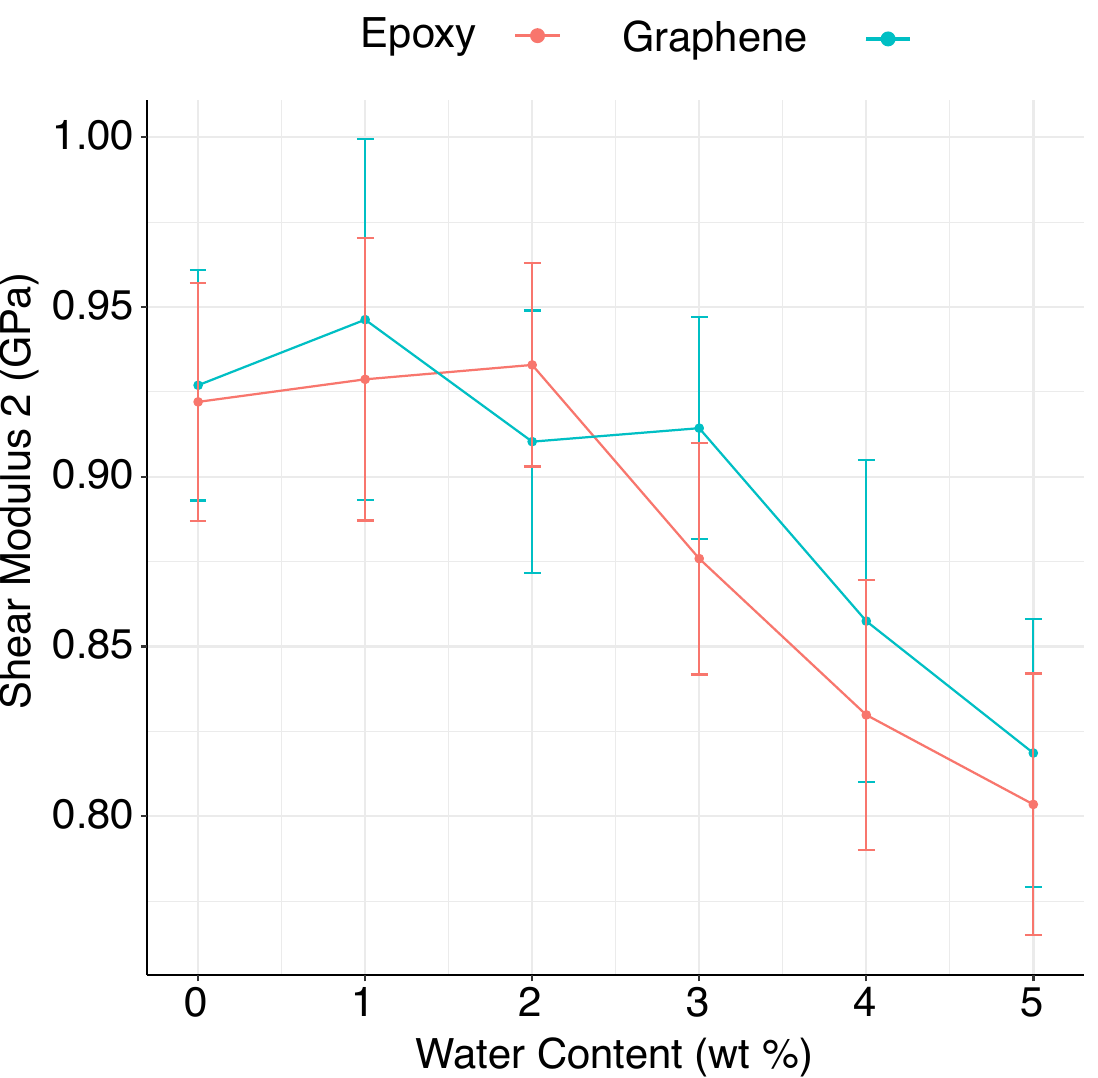}
  (e)
  \includegraphics[width=.25\textwidth]{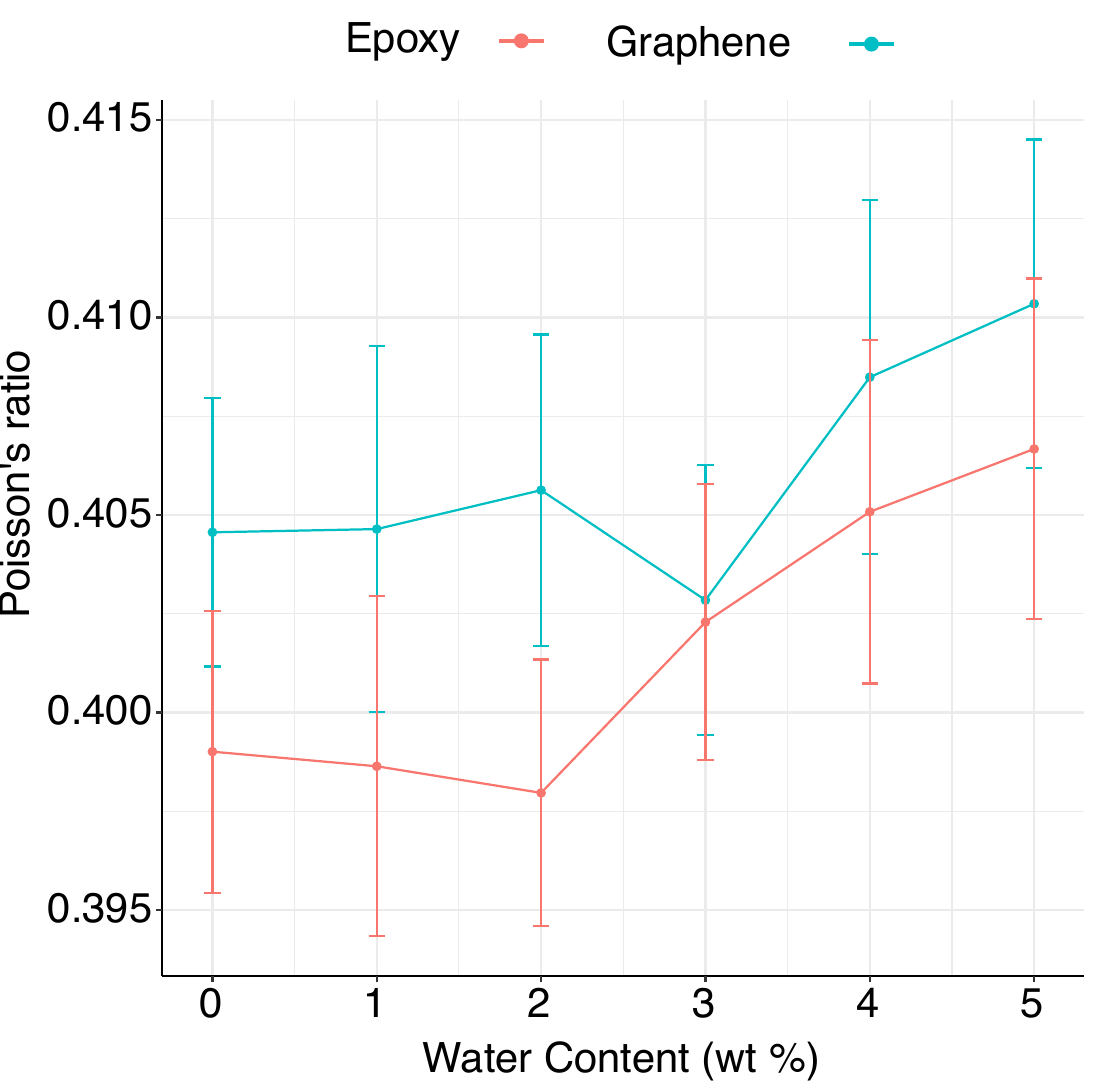}
  \caption{\textbf{Mechanical elastic properties.} Degradation of the elastic properties for both pure epoxy polymer and graphene nanocomposites, as a function of water content from $0$ \% to $5$ \% by mass. (a) Young's modulus, (b) bulk modulus, (c) and (d) shear moduli $G_1$ and $G_2$, and (e) Poisson's ratio. Error bars correspond to the total standard deviation of all $100$ replicas at a given water content and applied mechanical deformation.}\label{fig:elastic-water}
\end{figure}

The various elastic moduli introduced in section \ref{ssec:mech_methods} of the epoxy resin and the epoxy-graphene nanocomposite are computed as a function of the water content (see figure \ref{fig:elastic-water}. Our baseline prediction of the Young's modulus of the dry epoxy resin at approximately $2.6$ GPa agrees with experimental data.

The Young's modulus is similar for both systems, indicating minimal effect of the graphene sheets. As we discussed in our previous study, this is not unexpected as graphene sheets do not show significant enhancements to the Young's modulus until they have lateral dimensions of order $250$ nm, due to shear-lag effects and the lack of strong interactions between epoxy polymers and the graphene sheets \cite{suter_large_2023}. The shear modulus, largely governed by the behaviour of the epoxy matrix, remains similar between graphene and pure epoxy systems. However, we see significant differences in the bulk modulus and Poisson ratio with the addition of graphene. In the dry-state, the graphene-epoxy nanocomposite exhibits a higher bulk modulus than the neat epoxy. This arises because the graphene sheets act as incompressible inclusion that restrict volumetric deformation, increasing the effective bulk modulus. Overall, the relative behaviour of dry and hydrated systems remain consistent across the whole hydration range.

The evolution of mechanical properties with hydration show that both neat epoxy and graphene-epoxy remain stable up to $2$\% water content. Beyond this threshold, the water molecules are no longer primarily bound to the polar sites on the epoxy chains; the larger water clusters plasticizes the matrix reducing the Young's, bulk and shear moduli for both graphene containing and pure epoxy systems.
Inversely, the Poisson ratio increases above 2 \%wt water content, with the graphene system consistently higher. The increase in Poisson ratio for graphene inclusions can be rationalized via the relation: 
$
\nu = (3K - 2G) / (2(3K+G)) 
$
; the elevated $K$ relative to $G$ increases $\nu$. The relation $E=3K(1-2\nu)$ shows that the effects of higher $K$ and $\nu$ for graphene partially offset each other, leading to the Young's modulus remaining unchanged. \reviewsecond{Moreover, these independent observations of the various elastic properties confirm that the linear elastic isotropic behaviour of the material is preserved}.

%%%%%%%%%%%%%%%%%%%%%%%%%%%%%%%%%%%%%%%%%%%%%%%%%%%%%%%%%%%%%%%%%%%%%
%% Further discussions
%%%%%%%%%%%%%%%%%%%%%%%%%%%%%%%%%%%%%%%%%%%%%%%%%%%%%%%%%%%%%%%%%%%%%

\section{Uncertainty quantification in heterogeneous molecular models}

Polymers exhibit long relaxation times with respect to molecular dynamics accessible timescales, as such methods used for equilibration can lead to resulting molecular models with varying densities and structural configurations. To get a statistically representative sample of starting configurations, the starting structures of the mechanical testing simulations were the the result of an independent equilibration for over $100$ ns. Although we used unconventionally large ensembles of replica ($100$ replica) with very similar densities at $300$ K, one striking result from our mechanical simulations is the large variability of the mechanical properties predictions (see error bars in figure \ref{fig:elastic-water}). Therefore, it appeared essential to investigate in further detail the variability of our mechanical predictions across large ensembles. We now investigate the effect of ensemble size up to $1000$ replicas for the highly hydrated epoxy ($5$\% water content).

\begin{figure}[!ht]
    \centering
    \includegraphics[width=0.9\linewidth]{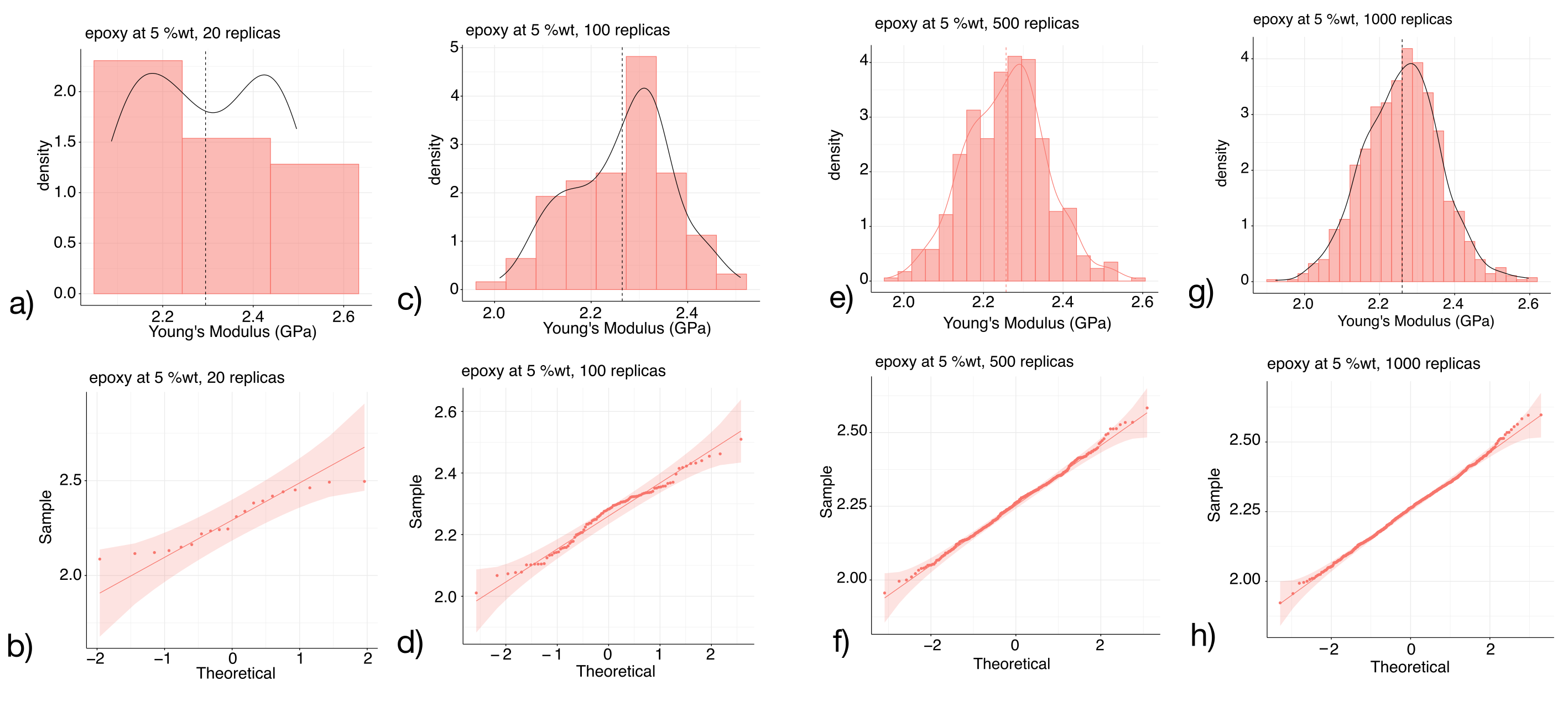}
    \caption{\textbf{Uncertainty quantification.} \reviewsecond{Analysis of the distribution of Young's modulus in pure epoxy resin with $5$ \% hydration using a non-bootstrapped histogram and a quantile-quantile plot against a theoretical Gaussian distribution for three ensemble size: (a, b) $20$ replicas, (c, d) $100$ replicas, and (e, f) $500$ replicas, and (g, h) $1000$ replicas. }} \label{fig:hists}
\end{figure}

We computed the Young's modulus for three ensemble sizes: $20$, $100$, $500$ and $1000$ replicas (see figure \ref{fig:hists}). The ensembles distributions are primarily described by means of histograms which the bin size is calculated using the Freedman–Diaconis rule~\cite{freedman_histogram_1981}. The histogram for $20$ replica appears to follow a bimodal distribution, with samples less likely to display the average Young's modulus of the ensemble at approximately $2.3$ GPa. Even for an ensemble of $100$ replica, the most probable value of the replicas Young's modulus exceeds the mean value of the ensemble. The distribution also appears skewed, biased toward higher values of the modulus. \reviewsecond{Only for $500$ and $1000$ replica ensembles, the distributions of the Young's modulus values start to resemble a Gaussian distribution, showing improved symmetry and centred around the mean value. Quantile-quantile plots of the distributions are also provided to assess the deviation from a Gaussian distribution. From these plots, the distributions are only clearly converged beyond $500$ replicas. The deviation from a Gaussian is only seen within extreme quantiles, below $-2$ and beyond $2$.}

\reviewsecond{In order to further assess, the influence of the number of the replicas on the shape of the distribution of the Young's modulus values, we computed the third and fourth moments of the distributions, that are the skewness and the excess kurtosis \label{fig:skewness_kurtosis}. These are indicators, respectively, of the asymmetry and the thickness of the tails of the distributions. Note that we focus on the excess kurtosis (kurtosis minus 3) to highlight the difference with a Gaussian distribution. The skewness of the distributions converge rapidly beyond $50$ replicas, independently of the presence of graphene. Similarly, the kurtosis of the distributions converge after $150$ replicas. The main difference in the distributions associated with the addition of graphene is the shift from positive skewness and kurtosis to negative skewness and kurtosis. Thus, the addition of graphene yields Young's modulus distributions biased toward values lower than the peak and thinner tails than a Gaussian distribution.}

\begin{figure}[!ht]
    \centering
    \includegraphics[width=0.9\linewidth]{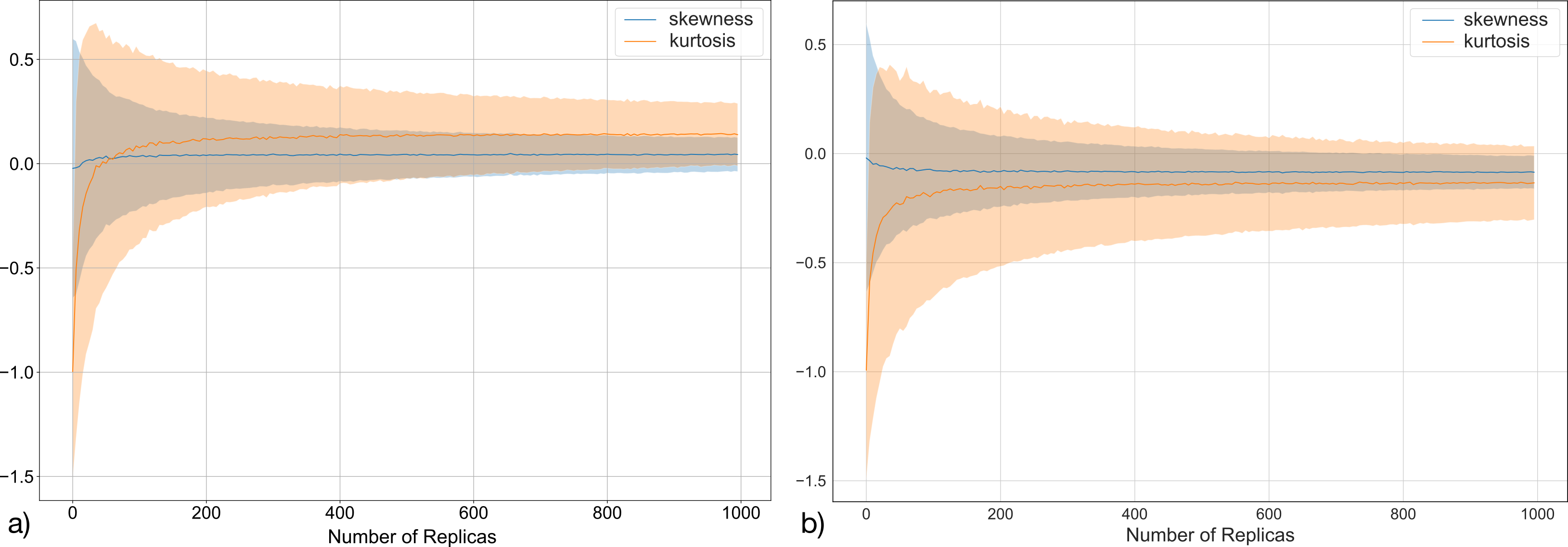}
    \caption{\reviewsecond{\textbf{Convergence of distributions.} Analysis of the evolution with the number of replica of the skewness and excess kurtosis of the distribution of the Young's modulus in (a) pure epoxy resin with $5$ \% hydration} and (b) epoxy-graphene nanocomposites with $5$ \% hydration. } \label{fig:kurtosis_skewness}
\end{figure}

Other factors not covered in this study may contribute to the uncertainty of the simulation. The two most notable are the interatomic force field of the system and the size of the system~\cite{edeling_global_2024}. The force field used was PCFF+, which has been shown to be significantly better than other classical all-atom force fields for modelling material properties such as the glass transition~\cite{bejagam_molecular_2020}. A previous study demonstrated that the Young's modulus of molecular models of pure epoxy resin only become independent of the model size when the simulation box lengths is above 40 \AA~\cite{wan_uncertainty_2021}. Thus our molecular models, of width approximately $100$ \AA, should not display size-induced variability of the mechanical properties.

%%%%%%%%%%%%%%%%%%%%%%%%%%%%%%%%%%%%%%%%%%%%%%%%%%%%%%%%%%%%%%%%%%%%%
%% Conclusions
%%%%%%%%%%%%%%%%%%%%%%%%%%%%%%%%%%%%%%%%%%%%%%%%%%%%%%%%%%%%%%%%%%%%%

\section{Conclusion}

The present study aimed at investigating in depth the influence of hydration on epoxy-based materials, and in particular their density, glass transition and mechanical properties. We have built large ensembles of molecular model of hydrated epoxy resins and epoxy-graphene nanocomposites. From these models, a campaign of MD simulations enables us to predict the systems’ \reviewsecond{thermomechanical} properties as well as the organisation of water molecular.

We have found that the main structural consequence of hydration was the growing size of water clusters in the resin and the nanocomposite. The addition of graphene reinforced the aggregation of the water molecules \reviewfirst{through hydrophobic effects}, with double the size of the largest water clusters produced in presence of graphene. \reviewfirst{Moreover, the use of oxidised graphene may help mitigate such clustering, instead favouring the dispersion of water molecules.}

Meanwhile, the short-range order of water molecules was neither perturbed by the increasing water content nor the insertion of graphene. We have also found that embedding graphene sheets inside an epoxy resin affects the bulk modulus and the Poisson ratio of the resulting nanocomposites. The average predicted moduli of the nanocomposite were systematically higher than that of the epoxy resin, independently of the water content. Conversely, the addition of graphene did not affect the glass transition temperature. Nonetheless, and consistently with structural observations, hydration diminishes the mechanical properties of resins and nanocomposites in a similar fashion. The effect of hydration on mechanical properties was particularly striking beyond \reviewsecond{$2$\% hydration. On the contrary, the effect of hydration on the glass transition temperature was found below $3$\% hydration.} Last, through uncertainty quantification of very large ensembles, we demonstrated that \reviewsecond{$150$} replicas are necessary to report converged distributions of elastic properties and, by extension, reproducible mechanical results.

Overall, our study emphasizes and clarifies the role of water molecules and hydration on the reduction of elastic properties of epoxy-based resins and nanocomposites. Our study also highlights the limited capability of graphene to mitigate the influence of hydration. \reviewsecond{Graphene-reinforced epoxy resins indeed display enhanced mechanical properties but suffer from identical softening on adsorption of water as does their pristine counterpart.}

\section{Acknowledgements}
This research was equally co-funded by the Engineering and Physical Sciences Research Council (EPSRC) and Hexcel. We acknowledge funding support from the European Union’s Horizon 2020 research and innovation program under Grant Agreement 800925 (VECMA project, www.vecma.eu), from the UK Engineering and Physical Sciences Research Council under Grant Agreements EP/W007711/1 (SEAVEA project, www.seavea-project.org), and the European Union’s Horizon 2020 Research and Innovation Programme under grant agreement 823712 (CompBioMed2, compbiomed.eu). We are grateful to the United States Department of Energy (DOE) via the Argonne Leadership Computing Facility for providing access to and node hours on Aurora under a DOE INCITE award in the years 2024-2025.

\section{Supplementary Information}
\reviewsecond{The Supplementary Information contains two figures: (i) the structure of water molecules around the graphene sheets in epoxy-graphene nanocomposites; and (ii) the evolution of the predictions of the Young's modulus of our epoxy resin models with the size of the molecular model and the duration of the deformation simulations.}

\bibliographystyle{elsarticle-num-names_perso} 
\bibliography{MyLibrary}

\end{document}